\documentclass[runningheads]{svmult}

\usepackage{makeidx}   
\usepackage{graphicx}  
\usepackage{subeqnar}  
\usepackage{multicol}  
\usepackage{physmubb}  
\makeindex             

\newcommand{\greeksym}[1]{{\usefont{U}{psy}{m}{n}#1}}
\newcommand{\umu}{\mbox{\greeksym{m}}}

\newcommand{\pom}{I\!\! P}

\begin{document}
\title*{Measurements of Diffractive Processes at CDF}
%
%
%
%
\titlerunning{Measurements of Diffractive Processes at CDF}
%
\author{Konstantiqn Goulianos (for the CDF Collaboration)
\protect\footnote{Presented at the 14$^{th}$ Topical Conference on Hadron Collider Physics,
HCP-2002, Karlsruhe, Germany, 29 Sep - 4 Oct 2002.}\\
The Rockefeller University, New York, NY 10021, USA}

%
\authorrunning{Konstantin Goulianos}
%
%

\maketitle              

\begin{abstract}
We review the results of measurements on hard diffractive processes  
performed by the CDF Collaboration
and report preliminary CDF results on two soft diffractive processes with 
a leading antiproton and a rapidity gap in addition to that 
associated with the antiproton. All results have been obtained from 
data collected in Run I of the Fermilab Tevatron $\bar pp$ collider.
\end{abstract}
\section{Introduction}
Diffractive $\bar pp$ interactions are characterized by a leading (high 
longitudinal momentum) outgoing proton or antiproton and/or a large  
{\em rapidity gap}, defined as a region of pseudorapidity, 
$\eta\equiv -\ln\tan\frac{\theta}{2}$, 
devoid of particles. The large rapidity gap is presumed to be due to the 
exchange of a Pomeron, which carries the internal quantum numbers of the 
vacuum. 
Rapidity gaps formed by multiplicity fluctuations in non-diffractive (ND) 
events are exponentially suppressed with increasing 
$\Delta\eta$, so that gaps of 
$\Delta\eta>3$ are mainly diffractive.
At high energies, where the available rapidity space is large, diffractive 
events may have more than one large gap. 

Diffractive events that incorporate a hard scattering are referred to as 
{\em hard diffraction}. 
In this paper we review briefly the results on hard diffraction published by
CDF and present preliminary results on two types of soft diffraction events 
with two diffractive rapidity gaps in an event, as shown schematically 
in Fig.~\ref{fig:SDD_DPE}.

\begin{minipage}[t]{0.5\textwidth}
\vglue -0.1cm
\includegraphics[width=1\textwidth]
{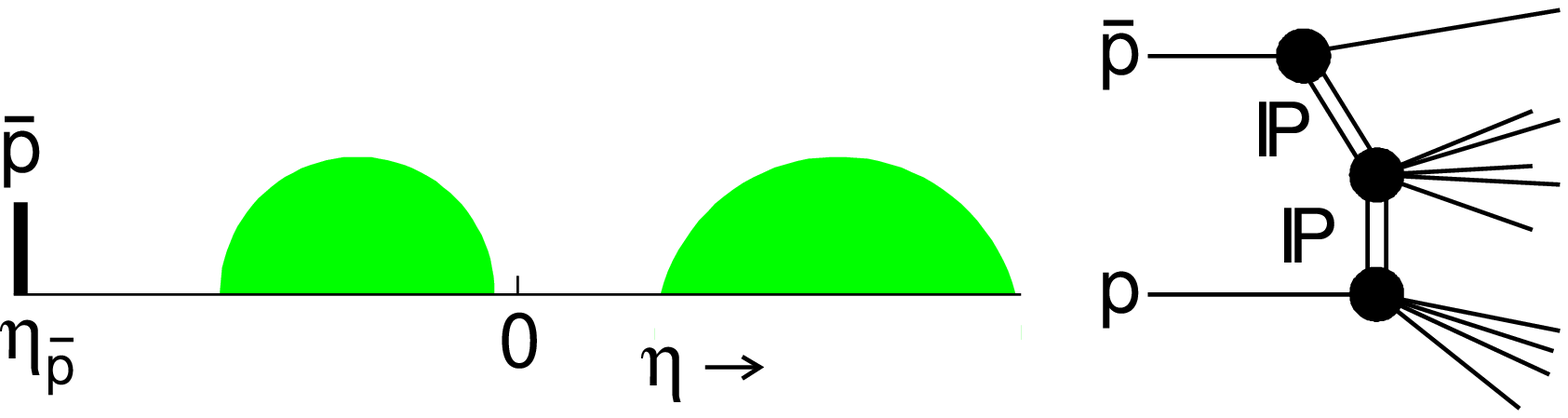}
\vglue -4.1cm
\centerline{{\large (a)} SDD}
\vglue 10cm
\end{minipage} 
\begin{minipage}[t]{0.5\textwidth} 
\vglue -0.4cm
\includegraphics[width=1\textwidth]
{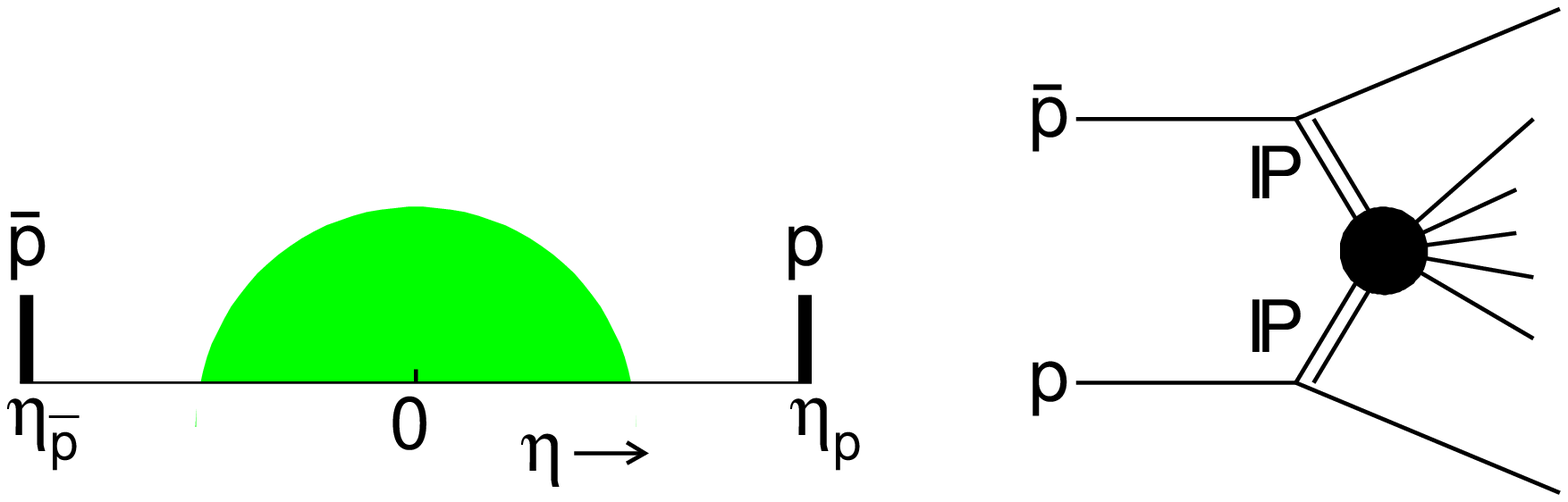}
\vglue -3.8cm
\centerline{{\large (b)} DPE}
\vglue -10cm
\end{minipage}
\begin{figure}
\vglue -1ex
\caption{Schematic drawings of pseudorapidity topologies and Pomeron 
exchange diagrams for (a) single plus double diffraction and (b) double 
Pomeron exchange.}
\label{fig:SDD_DPE}
\end{figure}
\vglue -3.8cm
\section{Hard diffraction}
The CDF results on hard diffraction fall into two classes, characterized 
by the signature used to identify and extract the diffractive signal:
a large rapidity gap or a leading antiproton. 
\subsection{Rapidity gap results}
Using the rapidity gap signature to identify diffractive events, 
CDF measured the single-diffractive (SD) 
fractions of $W$~\cite{CDF_W}, dijet~\cite{CDF_JJG}, $b$-quark~\cite{CDF_b} 
and $J/\psi$~\cite{CDF_jpsi} production in $\bar pp$ 
collisions at $\sqrt s=1800$ GeV and 
the fraction of dijet events with a rapidity gap between jets 
(double-diffraction - DD)
at $\sqrt s=1800$~\cite{CDF_JGJ1800} and 
630~\cite{CDF_JGJ630} Ge
V. The results for the measured fractions 
are shown in Table~\ref{table_frac}.

\begin{table}
\begin{center}
\caption{Diffractive fractions}
\begin{tabular}{lccc}
\hline
Hard process&$\sqrt{s}$ (GeV)&$R=\frac{\rm DIFF}{\rm TOTAL}\,(\%)$&
Kinematic region\\
\hline
{\fbox{SD}}&&&\\
$W(\rightarrow e\nu)$+G&1800&$1.15\pm 0.55$&$E_T^e,\;/\!\!\!\!E_T>20$ GeV\\
Jet+Jet+G&1800&$0.75\pm 0.1$&$E_T^{jet}>20$ GeV, $\eta^{jet}>1.8$\\
$b(\rightarrow e+X)$+G&1800&$0.62\pm 0.25$&$|\eta^e|<1.1$, $p_T^e>9.5$ GeV\\
$J/\psi(\rightarrow \mu\mu)$+G&1800&$1.45\pm 0.25$
&$|\eta^{\mu}|<0.6$, $p_T^{\mu}>2$ GeV\\
{\fbox{DD}}&&&\\
Jet-G-Jet&1800&$1.13\pm 0.16$&$E_T^{jet}>20$ GeV, $\eta^{jet}>1.8$\\
Jet-G-Jet&630&$2.7\pm 0.9$&$E_T^{jet}>8$ GeV, $\eta^{jet}>1.8$\\
\hline
\end{tabular}
\end{center}
\label{table_frac}
\end{table}

Since the different SD processes studied have different sensitivities 
to the gluon/quark ratio of the interacting partons, the approximate 
equality of the SD fractions at $\sqrt s=1800$ GeV 
indicates that the gluon fraction of the diffractive structure fraction
of the proton (gluon fraction of the Pomeron) is not very different 
from the proton's inclusive gluon fraction. By comparing the fractions 
of $W$, $JJ$ and $b$ production with Monte Carlo predictions, 
the gluon fraction of the Pomeron was found to be 
$f_g=0.54^{+0.16}_{-0.14}$~\cite{CDF_b}. 
This result was confirmed by a comparison of the diffractive 
structure functions obtained from studies of $J/\psi$ and $JJ$ production, 
which yielded a gluon fraction of $f_g^D=0.59\pm0.15$~\cite{CDF_jpsi}.

\subsection{Leading antiproton results}
Using a Roman pot spectrometer to detect leading antiprotons 
and determine their momentum and polar angle (hence the $t$-value),
CDF measured the ratio of SD to ND dijet production rates 
at $\sqrt s$=630~\cite{CDF_jj630} and 1800 GeV~\cite{CDF_jj1800} as a 
function of $x$-Bjorken of the struck parton in the $\bar p$. In leading order
QCD, this ratio is equal to the ratio of the corresponding 
structure functions. For dijet production, the relevant structure function is
the color-weighted combination of gluon and quark terms given by 
$F_{jj}(x)=x[g(x)+\frac49\sum_iq_i(x)]$. The diffractive structure function,
$\tilde{F}_{jj}^D(\beta)$, where $\beta=x/\xi$ is the momentum fraction of the 
Pomeron's struck parton, is obtained by multiplying 
the ratio of rates by the known $F_{jj}^{ND}$ and changing variables 
from $x$ to $\beta$ using $x\rightarrow \beta\xi$  
(the tilde over the $F$ indicates integration over 
$t$ and $\xi$). 

The CDF $\tilde{F}_{jj}^D(\beta)$ is presented in Fig.~\ref{fig:FD3}a 
and compared with 
a calculation based on diffractive parton densities obtained by the 
H1 Collaboration at HERA from a QCD fit to diffractive DIS data. The CDF 
result is suppressed by a factor of $\sim 10$ relative to the prediction from 
from HERA data, indicating a breakdown of factorization of approximately
the same magnitude as that observed in the rapidity gap data.

Factorization was also tested {\em within CDF data} by comparing the 
ratio of DPE/SD to that of SD/ND dijet production rates (Fig.~\ref{fig:FD3}b).
The DPE events were extracted from the leading antiproton 
data by requiring a rapidity gap in the forward detectors on the 
proton side. At $\langle \xi\rangle=0.02$ and 
$\langle x_{bj}\rangle =0.005$, the ratio of SD/ND to DPE/SD rates 
normalized per unit $\xi$ was found to be~\cite{CDF_DPE} $0.19\pm 0.07$,
violating factorization. 

\noindent\begin{minipage}[t]{0.5\textwidth}
\hspace{-0.5cm}\includegraphics[width=1\textwidth]
{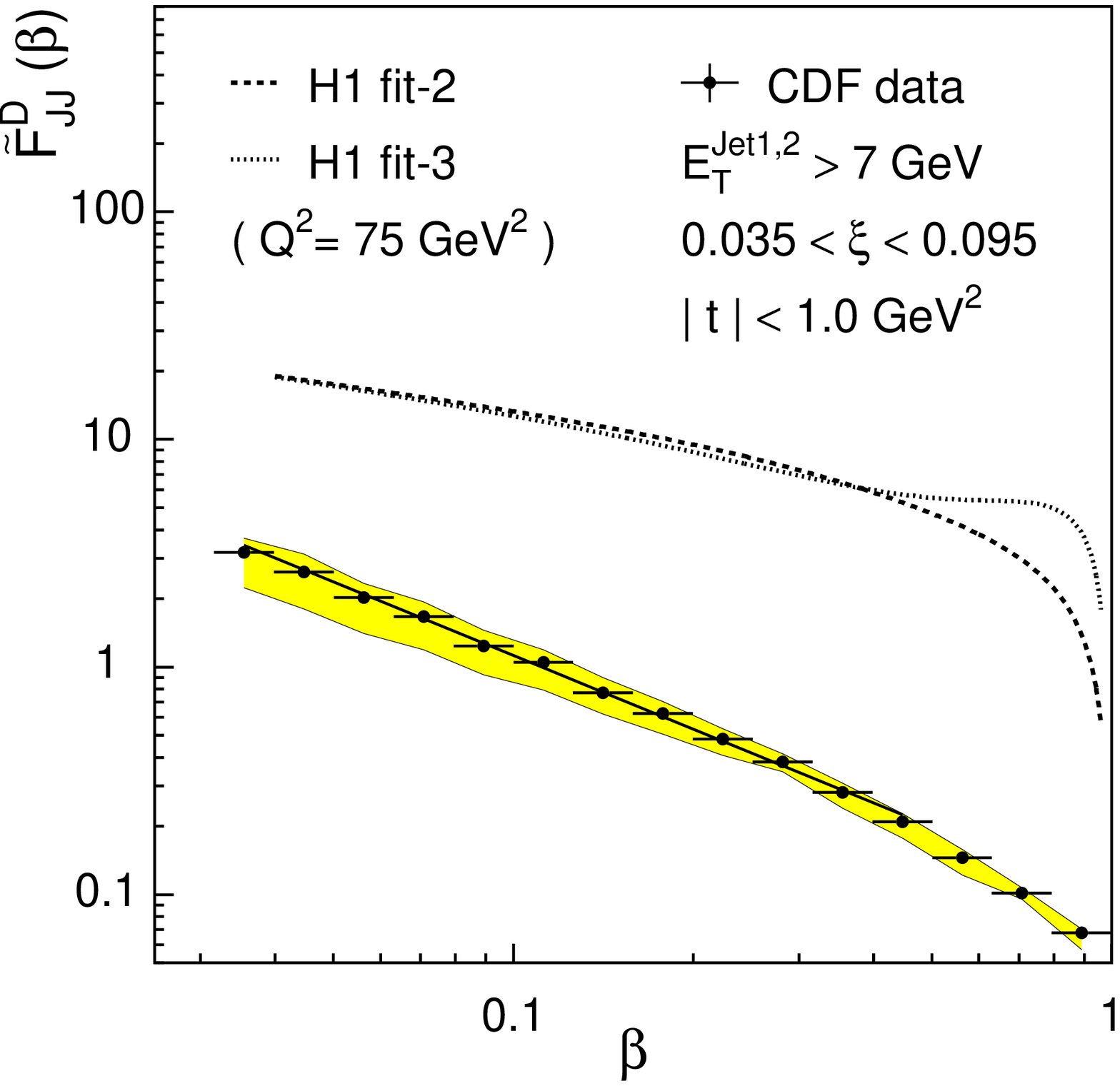}
\vspace*{-1.8cm}
\vglue 0.2cm
\vglue -0.2cm
\hspace{0.7\textwidth}{\LARGE (a)}
\end{minipage}
\begin{minipage}[t]{0.5\textwidth}
\includegraphics[width=1\textwidth]
{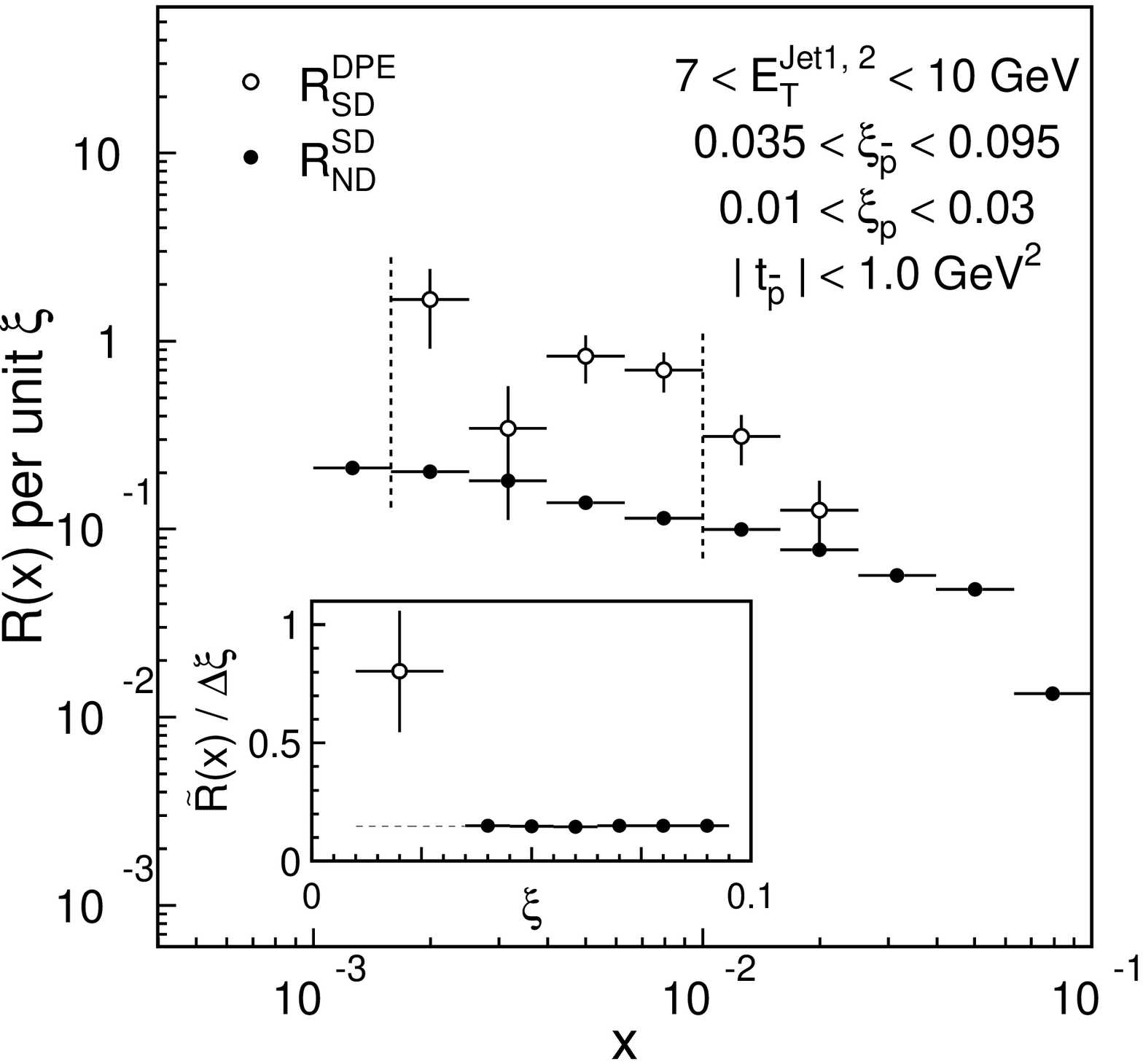}
\vspace*{-1.8cm}
\vglue -0.2cm
\vglue -0.2cm
\hspace{0.77\textwidth}{\LARGE (b)}
\end{minipage}
\begin{figure}
\vglue -1cm
\caption{(a)The diffractive structure function measured by CDF (data points 
and fit) compared with expectations based on the H1 fit 2 (dashed) and 
fit 3 (dotted) on diffractive DIS data at HERA (a more recent H1 fit 
on a more extensive data set yields a prediction similar in magnitude to 
that of fit 2 but with a shape which is in agreement with that of the 
CDF measurement).
(b) The ratio of DPE/SD rates compared with that of SD/ND rates as a function
of $x$-Bjorken of the struck parton in the escaping nucleon. The 
inequality of the two ratios indicates a breakdown of factorization. 
}
\label{fig:FD3}
\end{figure}

\section{Double-gap soft diffraction}
The motivation for studying events of the type shown in Fig.~\ref{fig:SDD_DPE} 
is its potential for providing further
understanding of the underlying mechanism responsible for 
the suppression of diffractive cross sections at high energies relative 
to Regge theory predictions. As shown in Fig.~\ref{fig:SD_DD}, 
such a suppression 
has been observed for both single diffraction (SD), 
$\bar p(p)+p\rightarrow [\bar p(p)+gap]+X$, and double diffraction (DD),
$\bar p(p)+p\rightarrow X_1+gap+X_2$.

\noindent\begin{minipage}[t]{0.5\textwidth}
\phantom{xxx}
\vspace*{-0.8cm}
\hspace{-0.5cm}\includegraphics[width=2.68in]
{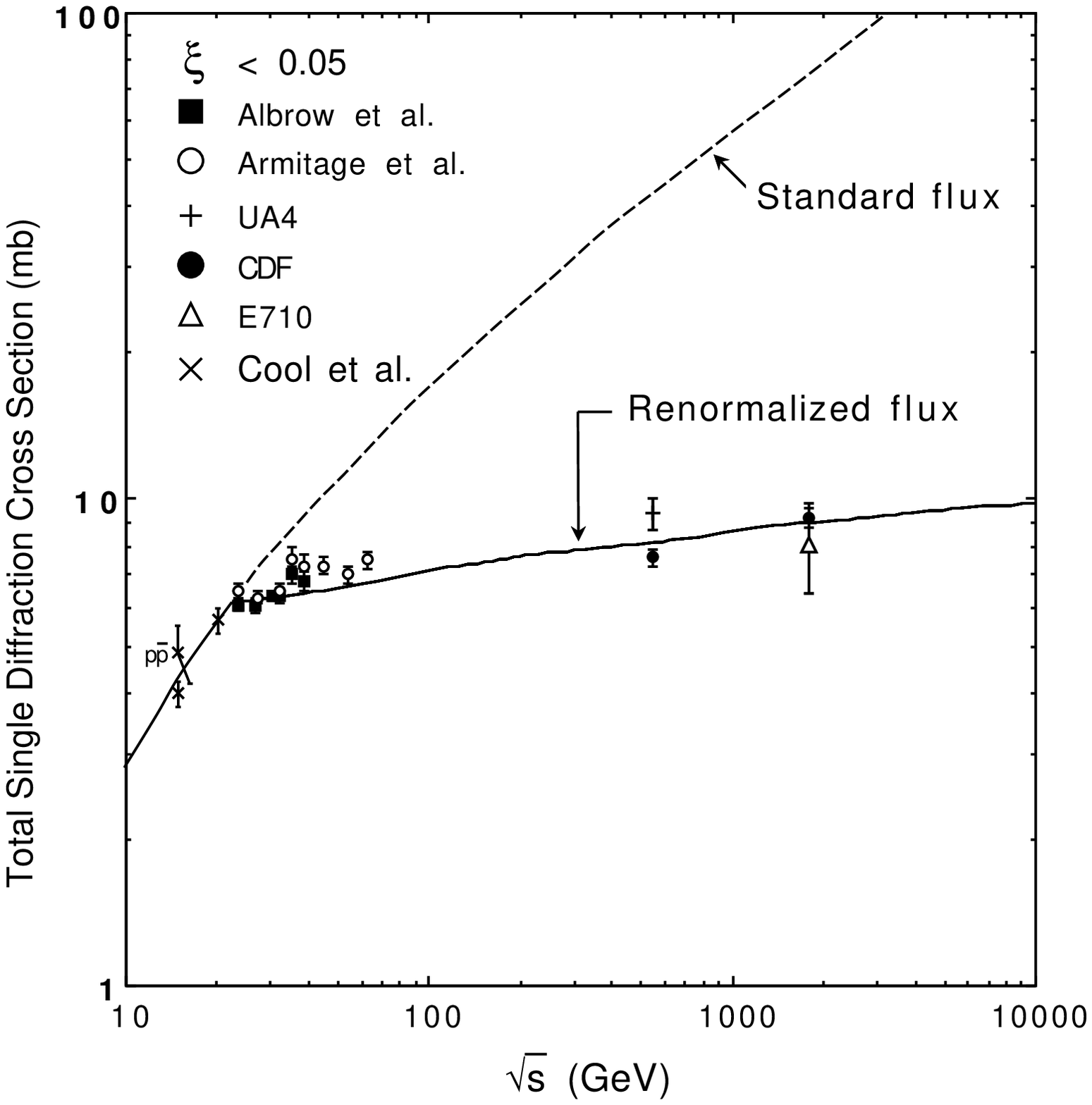}\vglue -2.05cm
\vglue -0.65in
\hspace{4.2cm}{\LARGE (a)}
\vglue 1.5in
\end{minipage}
\hspace*{0.8cm}
\begin{minipage}[t]{0.5\textwidth}
\phantom{xxx}
\hspace*{-1cm}\includegraphics[width=2.4in]
{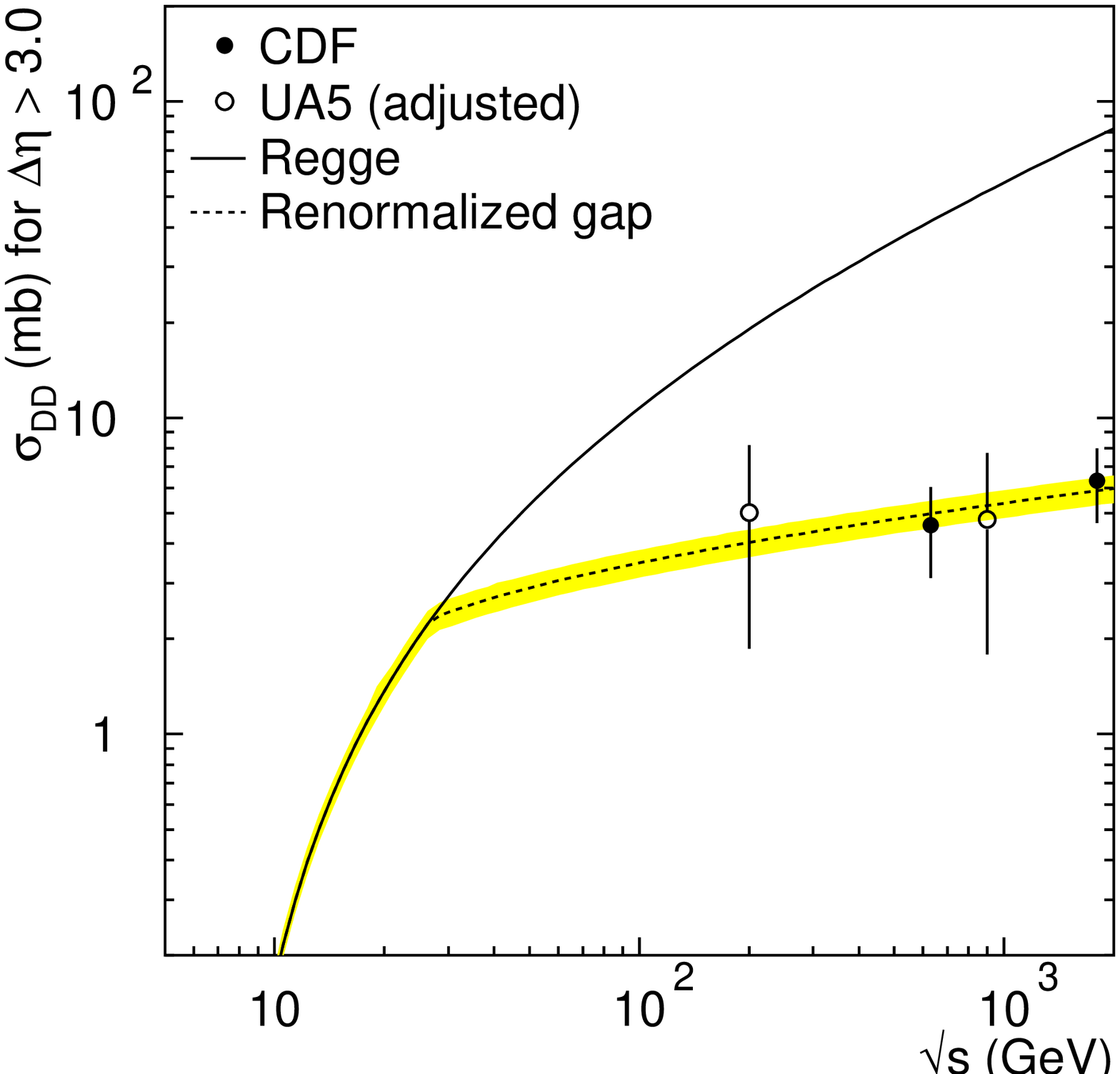}
\vglue 0.05cm
\vglue -0.65in
\hspace{3.6cm}{\LARGE (b)}
\vglue 1.5in
\end{minipage}
\vglue 1cm
\begin{figure}
\caption{(a) The $\bar pp$ total SD
cross section exhibits an $s$-dependence consistent
with the renormalization procedure of Ref.~\cite{R},
contrary to
the $s^{2\epsilon}$ behaviour expected from Regge theory
(figure from Ref.~\protect\cite{R});
(b) the $\bar pp$ total DD (central gap) cross section
agrees with the prediction of the
{\em renormalized rapidity gap} model~\cite{KGgap},
contrary to the $s^{2\epsilon}$
expectation from Regge theory
(figure from Ref.~\protect\cite{CDF_dd})}.
\label{fig:SD_DD}
\end{figure}

Naively, the suppression relative to Regge based predictions is
attributed to the spoiling of the diffractive 
rapidity gap by color exchanges in addition to Pomeron exchange.
In an event with two rapidity gaps, additional color exchanges would 
generally spoil both gaps. Hence, ratios of two-gap to one-gap 
rates should be unsuppressed. Measurements of such ratios could 
therefore be used to test the QCD aspects of gap formation without 
the complications arising from the rapidity gap survival probability.
 
\section{Data and results}
The data used for this study are inclusive SD event samples at $\sqrt s=1800$
and 630 GeV collected by triggering on a leading antiproton detected in 
a Roman Pot Spectrometer (RPS)~\cite{CDF_jj1800,CDF_jj630}. Below, we list the 
number of events used in each analysis within the indicated regions of 
antiproton fractional momentum loss $\xi_{\bar p}$ and 
4-momentum transfer squared $t$, 
after applying the vertex cuts $|z_{vtx}|<60$ cm and 
$N_{vtx}\le 1$ and a 4-momentum squared cut of $|t|<0.02$ GeV$^2$ (except for 
DPE at 1800 GeV for which $|t|<1.0$ GeV$^2$):

\begin{table}
\begin{center}
\caption{Events used in the double-gap analyses}
\begin{tabular}{lccc}
\hline
Process&$\xi$&Events at 1800 GeV&Events at 630 GeV\\
\hline
SDD&$0.06<\xi<0.09$&412K&162K\\
DPE&$0.035<\xi<0.095$&746K&136K\\
\hline
\end{tabular}
\end{center}
\end{table}

In the SDD analysis, the mean value of $\xi=0.07$ corresponds to a 
diffractive mass of $\approx 480\;(170)$ GeV at $\sqrt s=1800$ (630) GeV. 
The diffractive cluster X in such events covers almost the entire 
CDF calorimetry, which extends through the region $|\eta|<4.2$. 
Therefore, we use the same method of analysis as that used to extract the gap 
fraction in the case of DD~\cite{CDF_dd}. We search for {\em experimental gaps}
overlapping $\eta=0$, defined as regions 
of $\eta$ with no tracks or calorimeter
towers above thresholds chosen to minimize calorimeter noise contributions.
The results, corrected for triggering efficiency of $BBC_p$ (the beam 
counter array on the proton side) and converted to {\em nominal gaps} 
defined by $\Delta\eta=\ln\frac{s}{M_1^2M_2^2}$, are shown in 
Fig.~\ref{fig:SDD}.

\noindent\begin{minipage}[t]{0.5\textwidth}
\vspace*{0.1cm}
\includegraphics[width=1\textwidth]
{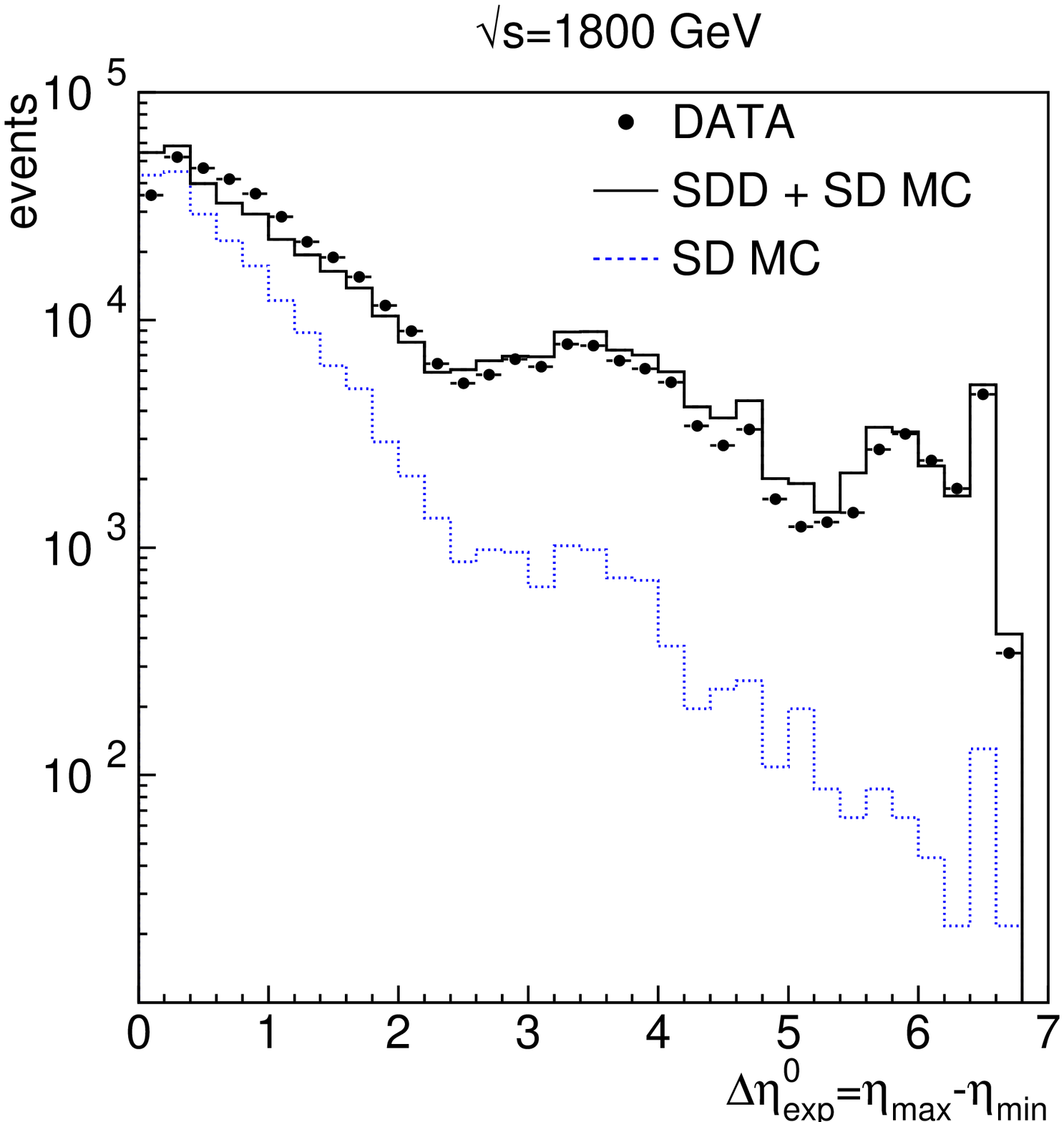}
\vspace*{0.2cm}
\vglue -2cm
\hspace{0.2\textwidth}{\LARGE (a)}
\end{minipage}
\begin{minipage}[t]{0.5\textwidth}
\vspace*{0.3cm}
\includegraphics[width=1\textwidth]
{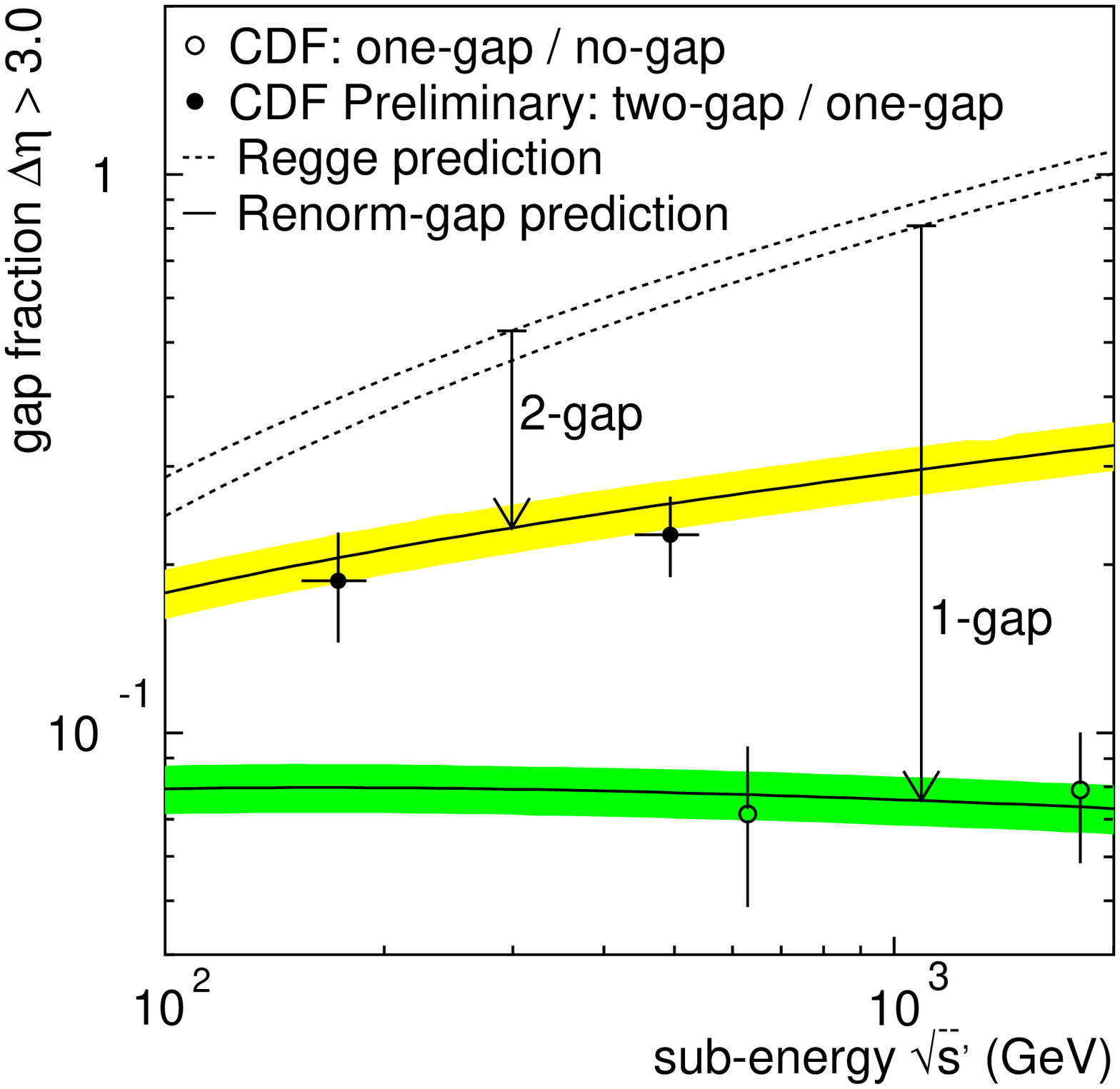}
\vspace*{-0.1cm}
\vglue -1.8cm
\hspace{0.2\textwidth}{\LARGE (b)}
\end{minipage}

\begin{figure}
\vglue 1ex
\caption{(a) The number of events as a function of
$\Delta\eta_{exp}^0=\eta_{max}-\eta_{min}$ for data at
$\protect\sqrt s=1800$ GeV (points), for SDD  Monte Carlo generated events
(solid line), and for only SD Monte Carlo events (dashed line);
(b) ratios of SDD to SD rates (points)
and DD to total (no-gap) rates (open circles) as a function
of $\sqrt{s'}$ of the sub-process $\pom p$  and of $\bar pp$,
respectively.
The uncertainties are highly correlated among all data points.}
\label{fig:SDD}
\end{figure}

The SDD Monte carlo simulation is based on Regge theory Pomeron exchange 
with the normalization left free to be determined from the data. 
The differential $dN/d\Delta\eta^0$ shape agrees with the theory 
(Fig.~\ref{fig:SDD}a), 
but the two-gap to one-gap ratio is suppressed (Fig.~\ref{fig:SDD}b). 
However, the 
suppression is not  as large as that in the one-gap to no-gap ratio. The bands 
through the data points represent predictions of the renormalized multigap 
parton model approach to diffraction~\cite{multigap}, 
which is a generalization of the renormalization models 
used for single~\cite{R} and double~\cite{KGgap} diffraction.   

In the DPE analysis, the $\xi_p$ is measured from calorimeter and 
beam counter information using the formula below and summing over 
all particles, defined experimentally as 
beam-beam counter (BBC) hits or calorimeter towers above 
$\eta$-dependent thresholds 
chosen to minimize noise contributions.
$$\xi^{\rm X}_p=\frac{M^2_{\rm X}}{\xi_{\bar p}\cdot s}=
\frac{\sum_i E_{\rm T}^i\;\exp(+\eta^i)}{\sqrt s}$$
\noindent For BBC hits we use the average value of $\eta$ of the BBC segment 
of the hit and an $E_T$ value 
randomly chosen from the expected $E_T$ distribution. 
The $\xi^X$ obtained by this method was calibrated by comparing 
$\xi^X_{\bar p}$, obtained by using $\exp(-\eta^i)$ in the above equation,
with the value of $\xi_{\bar p}^{RPS}$ measured by the Roman Pot Spectrometer.

Figure~\ref{fig:IDPE}a shows the $\xi^X_{\bar p}$ distribution 
for $\sqrt s=$1800 GeV.
The bump at $\xi_{\bar p}^X\sim 10^{-3}$ is attributed to
central calorimeter noise and is reproduced in Monte Carlo simulations. 
The variation of tower $E_T$ threshold across the various 
components of the CDF calorimetry does 
not affect appreciably the slope of the $\xi^X_{\bar p}$ distribution. 
The solid line represents the distribution measured in SD~\cite{CDF_PRD}.
The shapes of the DPE and SD distributions are in good agreement all the way 
down to the lowest values kinematically allowed.  

\begin{table}
\begin{center}
\caption{Double-gap to single-gap event ratios}
\begin{tabular}{lcc}
\hline
Source&$R^{DPE}_{SD}$(1800 GeV)&$R^{DPE}_{SD}$(630 GeV)\\
\hline
Data&$0.197\pm 0.010$&$0.168\pm 0.018$\\
$P_{gap}$ renormalization&$0.21\;\pm0.02$&$0.17\;\pm0.02$\\
Regge $\oplus$ factorization&$0.36\;\pm 0.04$&$0.25\;\pm 0.03$\\
$\pom$-flux renormalization&$0.041\pm 0.004$&$0.041\pm 0.004$\\
\hline
\end{tabular}
\end{center}
\label{table:twogap}
\end{table}

The ratio of DPE to inclusive SD events was evaluated for $\xi_p^X<0.02$.
The results for $\sqrt s=$1800 and 630 GeV are presented in 
Table~\ref{table:twogap} 
and shown in Fig.~\ref{fig:IDPE}b. Also presented in the table are the 
expectations from gap probability renormalization~\cite{multigap}, 
Regge theory and factorization, and Pomeron flux renormalization 
for both exchanged Pomerons~\cite{R}. The  quoted uncertainties
are largely systematic for both data and theory; the theoretical 
uncertainties of 10\% are due to the uncertainty in the ratio of the 
triple-Pomeron to the Pomeron-nucleon couplings~\cite{GM}. 

The data are in excellent agreement with the predictions of the gap 
renormalization approach.

\noindent\begin{minipage}[t]{0.5\textwidth}
\vspace*{1.5cm}
\hspace{-0.5cm}\includegraphics[width=1\textwidth]
{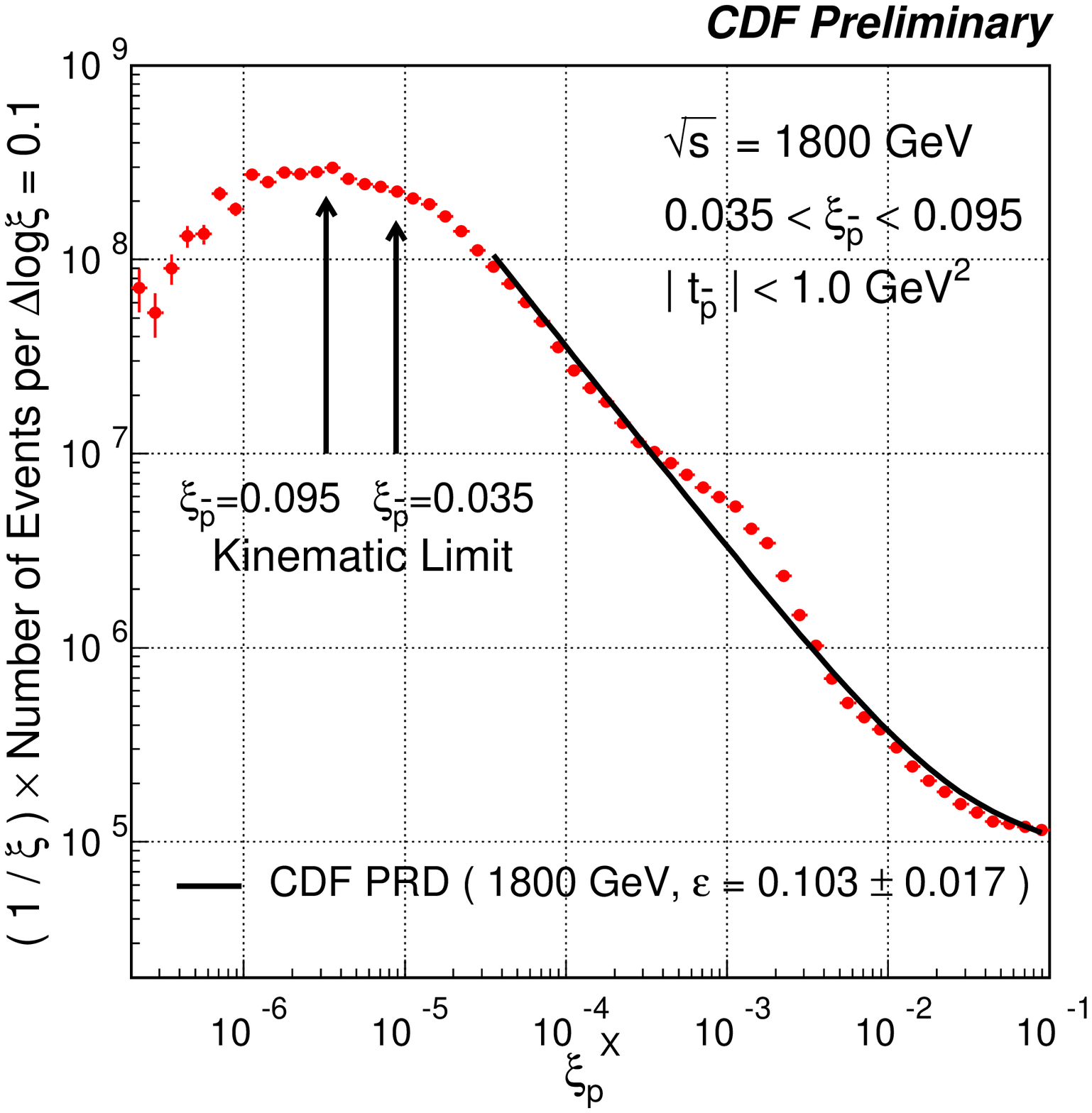}
\vspace*{-1.8cm}
\vglue 0.2cm
\vglue -3.2cm
\hspace{0.7\textwidth}{\LARGE (a)}
\end{minipage}
\begin{minipage}[t]{0.5\textwidth}
\vspace*{1.5cm}
\includegraphics[width=1\textwidth]{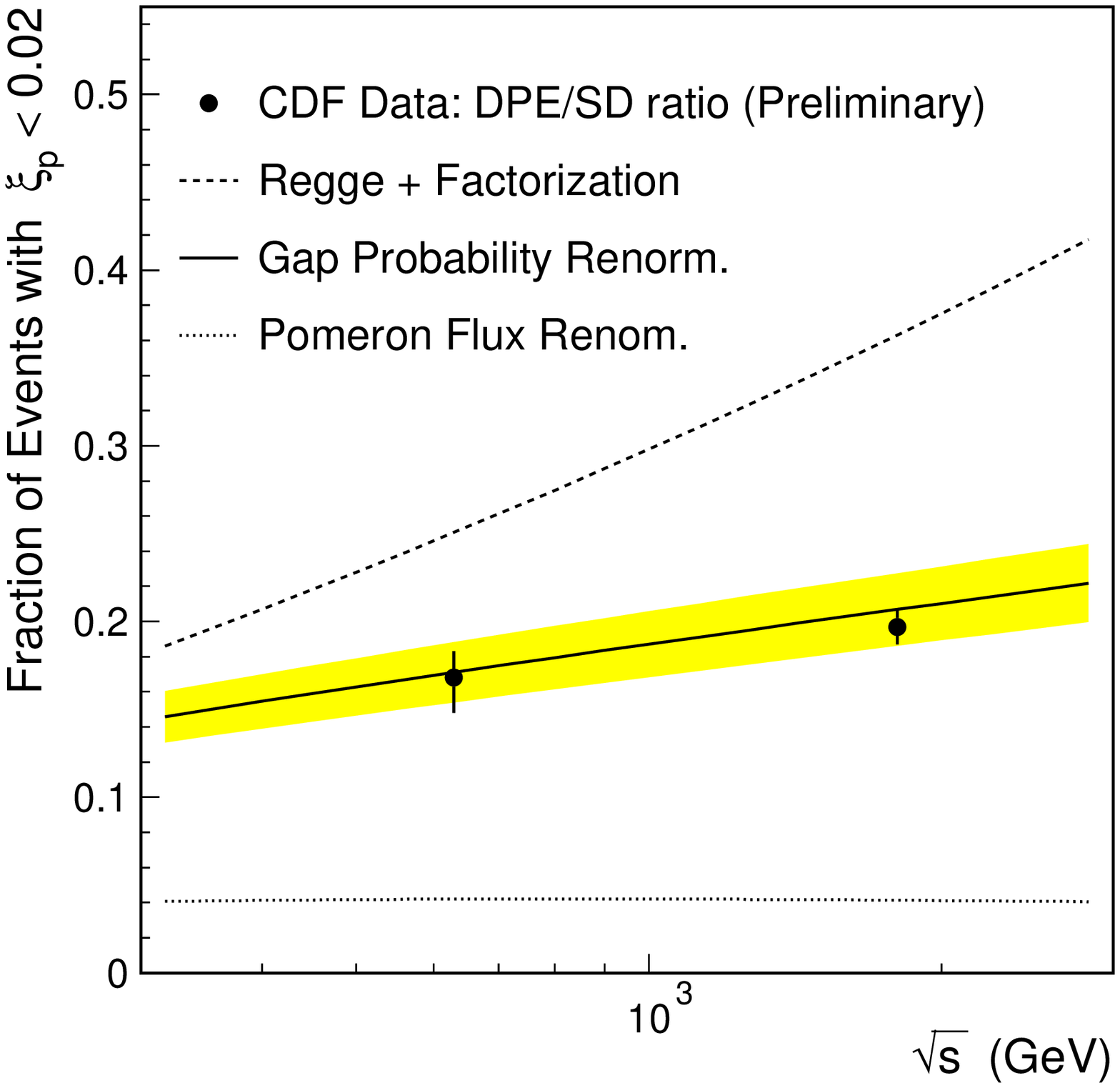}
\vspace*{-1.8cm}
\vglue -0.2cm
\vglue -3.2cm
\hspace{0.77\textwidth}{\LARGE (b)}
\end{minipage}
\begin{figure}
\vglue 2cm
\caption{(a) $\xi_{\bar p}^X$ distribution at $\sqrt s=$1800 GeV 
for events with a $\bar p$ of 
$0.035<\xi_{\bar p}^{RPS}<0.095$. 
The solid line is the distribution obtained in single
diffraction dissociation. The bump at $\xi_{\bar p}^X\sim 10^{-3}$ is due to 
central calorimeter noise and is reproduced in Monte Carlo simulations. 
(b)Measured ratios of DPE to SD rates (points)
compared with predictions based on Regge theory(dashed), Pomeron flux 
renormalization for both exchanged Pomerons (dotted) and gap probability 
renormalization (solid line).}
\label{fig:IDPE}
\end{figure}

\end{document}